\pdfoutput=1

\documentclass[11pt,twoside,a4paper,cmspaper,final,collab]{cms-tdr}
\def\svnVersion{481714P}\def\cmsCernNoTag{CERN-EP-2018-078}\def\cmsCernDate{\today}\def\cmsMessage{Published in the Journal of High Energy Physics as \href{http://dx.doi.org/10.1007/JHEP11(2018)018}{\doi{10.1007/JHEP11(2018)018.}}}

\begin{document}\cmsNoteHeader{HIG-17-029}

\hyphenation{had-ron-i-za-tion}
\hyphenation{cal-or-i-me-ter}
\hyphenation{de-vices}
\RCS$HeadURL: svn+ssh://svn.cern.ch/reps/tdr2/papers/HIG-17-029/trunk/HIG-17-029.tex $
\RCS$Id: HIG-17-029.tex 472110 2018-08-17 12:06:19Z ccaillol $

\newlength\cmsTabSkip\setlength{\cmsTabSkip}{1ex}
\providecommand{\CL}{CL\xspace}
\newcommand{\ma}{\ensuremath{m_{\Pa}}\xspace}
\newcommand{\qqZZ}{\ensuremath{\qqbar\to\cPZ\cPZ}\xspace}
\newcommand{\ggZZ}{\ensuremath{\Pg\Pg\to\cPZ\cPZ}\xspace}
\newcommand{\processmmtt}{\ensuremath{{\Ph}\to{\Pa\Pa}\to2\PGm2{\Pgt}}\xspace}
\newcommand{\processtttt}{\ensuremath{{\Ph}\to{\Pa\Pa}\to4{\Pgt}}\xspace}

\cmsNoteHeader{HIG-17-029}

\title{Search for an exotic decay of the Higgs boson to a pair of light pseudoscalars in the final state of two muons and two \texorpdfstring{$\Pgt$}{tau} leptons in proton-proton collisions at $\sqrt{s}=13\TeV$}

\date{\today}

\abstract{
A search for exotic Higgs boson decays to light pseudoscalars in the final state of two muons and two $\Pgt$ leptons is performed using proton-proton collision data recorded by the CMS experiment at the LHC at a center-of-mass energy of 13\TeV in 2016, corresponding to an integrated luminosity of 35.9\fbinv. Masses of the pseudoscalar boson between 15.0 and 62.5\GeV are probed, and no significant excess of data is observed above the prediction of the standard model. Upper limits are set on the branching fraction of the Higgs boson to two light pseudoscalar bosons in different types of two-Higgs-doublet models extended with a complex scalar singlet.
}

\hypersetup{%
pdfauthor={CMS Collaboration},%
pdftitle={Search for an exotic decay of the Higgs boson to a pair of light pseudoscalars in the final state of two muons and two tau leptons in proton-proton collisions at sqrt(s)=13 TeV},%
pdfsubject={CMS},%
pdfkeywords={CMS, physics, Higgs boson, exotic decays, NMSSM, 2HDM+S}}

\maketitle

\section{Introduction}
\label{sec:intro}

In 2012 the ATLAS and CMS Collaborations discovered a particle with a mass of 125\GeV~\cite{Aad:2012tfa,Chatrchyan:2012xdj,Chatrchyan:2013lba} compatible with the Higgs boson predicted in the standard model (SM) of particle physics~\cite{Englert:1964et,Higgs:1964ia,Higgs:1964pj,Guralnik:1964eu,Higgs:1966ev,Kibble:1967sv}. Although all the measurements of the couplings and properties of this particle indicate compatibility with the SM within the experimental uncertainties, the existence of exotic decays of the Higgs boson is still allowed. The combination of data collected at center-of-mass energies of 7 and 8\TeV by ATLAS and CMS constrains branching fractions of the Higgs boson to particles beyond the SM to less than 34\% at 95\% confidence level (\CL)~\cite{Khachatryan:2016vau}.

Many well-motivated exotic decays of the Higgs boson are proposed in theories beyond the SM~\cite{PhysRevD.90.075004}. A possible scenario consists of exotic Higgs boson decays to pairs of light pseudoscalars, which subsequently decay to pairs of SM particles. Such a process would be allowed in two-Higgs-doublet models (2HDM) extended with a scalar singlet (2HDM+S)~\cite{PhysRevD.90.075004}. In 2HDM+S, 5 scalar and 2 pseudoscalar particles are  predicted: one of the scalars, $\Ph$, can be compatible with the discovered Higgs boson, while one of the pseudoscalars, $\Pa$, can be light enough so that $\Ph\to \Pa\Pa$ decays are allowed. The next-to-minimal supersymmetric SM (NMSSM) is a particular case of 2HDM+S~\cite{Ellwanger:2009dp,Maniatis:2009re}.

The ATLAS and CMS Collaborations have set limits on exotic decays of the Higgs boson to a pair of light pseudoscalar bosons, in different final states and in various ranges of the pseudoscalar mass, $\ma$~\cite{Khachatryan:2017mnf,Khachatryan:2015nba,Khachatryan:2015wka,Aaboud:2016oyb,Aad:2015sva,Aad:2015bua,Aad:2015oqa}. In particular, CMS published a null result in the search in the $2\Pgm 2\Pgt$ final state for $15.0<\ma<62.5\GeV$ using data collected at a center-of-mass energy of 8\TeV~\cite{Khachatryan:2017mnf}, and ATLAS reported a null result in the same final state at the same energy for $3.7<\ma<50.0\GeV$ using special reconstruction techniques for Lorentz-boosted $\Pgt$ lepton pairs~\cite{Aad:2015oqa}.

This paper presents a search for an exotic decay of the Higgs boson to a pair of light pseudoscalar bosons in the final state of two muons and two $\Pgt$ leptons. The analysis is based on data collected in 2016 by the CMS experiment in proton-proton (pp) collisions at a center-of-mass energy of 13\TeV, corresponding to an integrated luminosity of 35.9\fbinv. Masses of the pseudoscalar boson between 15.0 and 62.5\GeV are probed. Below 15\GeV, the pseudoscalar bosons are Lorentz-boosted, causing their decay products to be collimated and to fail the isolation selection criteria used in this analysis.  The analysis scans the reconstructed dimuon mass spectrum for a characteristic resonance structure. Four different final states are studied to cover the different possible $\Pgt$ lepton decay modes: $\Pgm\Pgm +\Pe\Pgm$, $\Pgm\Pgm +\Pe\tauh$, $\Pgm\Pgm +\Pgm\tauh$, and $\Pgm\Pgm +\tauh\tauh$,  where $\tauh$ denotes a $\Pgt$ lepton decaying hadronically. The $\Pgm\Pgm +\Pe\Pe$ and $\Pgm\Pgm +\Pgm \Pgm$ final states are not considered because of their smaller branching fractions and the large background contribution from $\PZ$ boson pair production.
The event selection and signal extraction used in this analysis have been optimized for the $\processmmtt$ decay channel, where $\Ph$ has a mass of 125\GeV. Events from the $\processtttt$ process can also enter the signal region when at least two of the $\Pgt$ leptons decay leptonically to muons and neutrinos. These events are treated as a part of the signal even if they do not exhibit a narrow dimuon mass peak. Assuming 2HDM-like scenarios, the ratio of the branching fractions of $\Pa\to2\Pgm$ and $\Pa\to2\Pgt$ is proportional to the ratio of the squared masses of the muon and the $\Pgt$ lepton:
\begin{linenomath}
\begin{equation}\label{eq:mm}
\frac{\mathcal{B}(\Pa\to2\Pgm)}{\mathcal{B}(\Pa\to2\Pgt)}=\frac{m_{\Pgm}^2\sqrt{1-(2m_{\Pgm}/\ma)^2}}{m_{\PGt}^2\sqrt{1-(2m_{\PGt}/\ma)^2}}\simeq \frac{m_{\Pgm}^2}{m_{\PGt}^2}.
\end{equation}
\end{linenomath}

Events are selected only if the invariant mass of the four objects in the final state is below 100--130\GeV (depending on the final state) to enforce the compatibility with a Higgs boson decay. This criterion strongly suppresses both the background from events with genuine leptons, which arise mostly from the $\PZ$ boson pair production, and the backgrounds with jets misidentified as $\Pgt$ leptons, leaving only a few expected background events in the signal region. The background from $\PZ$ boson pair production is estimated from simulation, whereas the background with jets misidentified as $\Pgt$ leptons is estimated from data, as detailed in Section~\ref{sec:bkg}. The presence of a signal is probed using the reconstructed dimuon mass as an observable. Given the narrow width of the signal and the small number of expected background events, signal and background distributions are parameterized to perform an unbinned maximum-likelihood fit.

\section{The CMS detector}

The central feature of the CMS apparatus is a superconducting solenoid of 6\unit{m} internal diameter, providing a magnetic field of 3.8\unit{T}. Within the solenoid volume, there are a silicon pixel and strip tracker, a lead tungstate crystal electromagnetic calorimeter (ECAL), and a brass and scintillator hadron calorimeter, each composed of a barrel and two endcap sections. Forward calorimeters extend the pseudorapidity coverage provided by the barrel and endcap detectors. Muons are detected in gas-ionization chambers embedded in the steel flux-return yoke outside the solenoid. Events of interest are selected using a two-tiered trigger system~\cite{Khachatryan:2016bia}. A more detailed description of the CMS detector, together with a definition of the coordinate system used and the relevant kinematic variables, can be found in Ref.~\cite{Chatrchyan:2008zzk}.

\section{Simulated samples and event reconstruction}

Signal processes, for both $\processmmtt$ and $\processtttt$, are generated using the \MGvATNLO 2.2.2 generator~\cite{Alwall:2014hca} with its implementation of the 2HDM and the NMSSM, in gluon fusion and vector boson fusion production. They are simulated at leading order (LO) in perturbative quantum chromodynamics (QCD) with the MLM jet matching and merging scheme~\cite{Alwall:2007fs}.
The generator is interfaced with \PYTHIA 8.212~\cite{Sjostrand:2014zea} to model the parton showering and fragmentation as well as the decay of the $\Pgt$ leptons. The {CUETP8M1} tune~\cite{Khachatryan:2015pea} is chosen for the \PYTHIA parameters controlling the description of the underlying event. The $\PZ\PZ$ background from quark-antiquark annihilation is generated at next-to-LO (NLO) in perturbative QCD with \POWHEG v2.0~\cite{Alioli:2008gx,Nason:2004rx,Frixione:2007vw}, while the $\ggZZ$ process is generated at LO with \MCFM 7.0~\cite{Campbell:2010ff}. The set of parton distribution functions is NLO NNPDF3.0 for NLO samples, and LO NNPDF3.0 for LO samples~\cite{Ball:2014uwa}. The fully differential cross section for the {\qqZZ} process has been computed at next-to-NLO (NNLO)~\cite{Grazzini2015407}, and the NNLO/NLO $K$-factor is applied to the \POWHEG sample as a function of the invariant mass of the $\PZ$ boson pair. Rare processes, such as triboson, $\ttbar\PZ$, or SM Higgs boson production, have a negligible contribution to the signal region because they typically have a larger invariant mass of the four leptons in the final state.

Simulated samples include additional $\Pp\Pp$ interactions per bunch
crossing (pileup), and are reweighted so as to match the pileup distribution observed in data. Generated events are processed through a simulation of the CMS detector based on
\GEANTfour~\cite{Agostinelli:2002hh}.

The reconstruction of events relies on the particle-flow (PF) algorithm~\cite{Sirunyan:2017ulk},
which combines the information from the CMS subdetectors to identify
and reconstruct the particles emerging from $\Pp\Pp$ collisions:
charged and neutral hadrons, photons, muons, and electrons.
Combinations of these PF objects are used to reconstruct
higher-level objects such as jets or $\tauh$ candidates.
The reconstructed vertex with the largest value of summed physics-object $\pt^2$ is taken to be the primary $\Pp\Pp$ interaction vertex, where $\pt$ denotes the transverse momentum. The physics objects are the jets, clustered using a jet-finding algorithm~\cite{Cacciari:2008gp,Cacciari:2011ma} with the tracks assigned to the vertex as inputs, and the associated missing transverse momentum, taken as the negative vector sum of the \pt of those jets.

Electrons are reconstructed by matching ECAL clusters to tracks in the tracker.  They are then identified with a multivariate discriminant that makes use of variables related to energy deposits in the ECAL, to the quality of the track, and to the compatibility between the ECAL clusters and the track that have been matched together~\cite{Khachatryan:2015hwa}. Muons are reconstructed by building tracks from hits in the tracker and in the muon system, and are identified using variables related to the number of measurements in the tracker and the muon systems and to the quality of the track reconstruction~\cite{Chatrchyan:2012xi}. They are required to have a relative isolation less than 0.2, with the relative isolation variable defined as follows:
\begin{linenomath}
\begin{equation}
I^{\Pgm} \equiv \frac{\sum_\text{charged}  \PT + \max\left( 0, \sum_\text{neutral}  \PT
                                         - \frac{1}{2} \sum_\text{charged, PU} \PT  \right )}{\PT^{\Pgm}}.
\label{eq:reconstruction_isolation}
\end{equation}
\end{linenomath}
In this equation, $\sum_\text{charged}  \PT$ is the scalar \pt sum of the charged particles associated with
the primary vertex in a cone of size
$\Delta R = \sqrt{\smash[b]{(\Delta \eta)^2 + (\Delta \phi)^2}} = 0.4$ around the muon direction. The sum
$\sum_\text{neutral}  \PT$  is
a similar quantity for neutral particles. The \pt of neutral particles originating from pileup vertices is considered on the basis of simulation to be half of that of charged particles associated with pileup vertices, denoted by $\sum_\text{charged, PU} \PT$. The term $\PT^{\Pgm}$ denotes the muon \pt. The azimuthal angle, $\phi$, is expressed in radians.

Jets are reconstructed from PF objects with the anti-\kt clustering algorithm implemented
in the \FASTJET library~\cite{Cacciari:2011ma, Cacciari:fastjet2}, using a distance parameter of 0.4. Jets that originate from {\cPqb} quarks, called {\cPqb} jets, are identified with the combined secondary vertex (CSVv2) algorithm~\cite{Sirunyan:2017ezt}. The algorithm builds a discriminant from variables related to potential secondary vertices associated to the jet, and from track-based lifetime information. The working point chosen in this search provides an efficiency for {\cPqb} quark jets of approximately 70\%, and a misidentification rate for light-flavor jets of approximately 1\%. Events with reconstructed {\cPqb} jets with $\pt>20\GeV$ are vetoed in this analysis to reject $\ttbar$ events and other backgrounds with {\cPqb} quark jets.

Hadronically decaying $\Pgt$ leptons are reconstructed with the hadrons-plus-strips algorithm~\cite{Khachatryan:2015dfa, CMS-PAS-TAU-16-002}. This algorithm starts from anti-\kt jets and reconstructs $\tauh$ candidates from tracks and energy deposits in strips of the ECAL, in the 1-prong, 1-prong + $\pi^0$, 2-prong, and 3-prong decay modes. The 2-prong decay mode allows $\tauh$ candidates to be reconstructed even if one track has not been reconstructed. Given the large rate for jets to be misidentified in this decay mode and the limited increase in efficiency for genuine $\tauh$ candidates, the 2-prong decay mode is not used to reconstruct $\tauh$ candidates in the signal region of this analysis, but is used in some control regions to study events with jets misidentified as $\tauh$ candidates. Hadronically decaying $\Pgt$ leptons are further required to be identified using a multivariate discriminator that combines isolation and lifetime variables. The working point of the discriminator has a $\tauh$ identification efficiency of approximately 57\% for a misidentification rate of light-flavor jets of approximately 0.35\%. Discriminators to reject muons and electrons misidentified as $\tauh$ candidates are further applied.

\section{Event selection}

Online, events are required to pass a double-muon trigger with \pt thresholds of 17 and 8\GeV for the leading and subleading muons, respectively, or a single-muon trigger with a \pt threshold of 24\GeV. In the $\Pgm\Pgm+\Pe\Pgm$ and $\Pgm\Pgm+\Pgm\tauh$ final states, events are also selected if they pass a triple-muon trigger with \pt thresholds of 12, 10, and 5\GeV. Offline, the leading muon must have $\pt>18\GeV$ (or 25\GeV if only the single-muon trigger is satisfied), and the subleading one $\pt>9\GeV$ (or 11\GeV if only the triple-muon trigger is satisfied). Selecting muons offline with \pt thresholds 1\GeV above the online thresholds ensures fully efficient triggers in this analysis. If there are additional muons, each is required to have $\pt>5\GeV$ (or 6\GeV if only the triple-muon trigger conditions have been met). All muons must satisfy $\abs{\eta}<2.4$. Electrons from $\Pgt$ lepton decays are required to have $\pt>7\GeV$ and $\abs{\eta}<2.5$, and $\tauh$ candidates are required to satisfy $\pt>18.5\GeV$ and $\abs{\eta}<2.3$.  Each event is required to have an opposite-sign (OS) pair of isolated muons and an OS pair of isolated $\Pgt$ candidates ($\Pe$, $\Pgm$, or $\tauh$).

In final states with three muons, the highest \pt muon is considered as originating promptly from the decay of the pseudoscalar bosons. It is paired with the next-highest \pt  OS muon. The third muon is considered as a decay product of a $\Pgt$ lepton. The probability for success of this algorithm for the expected signal varies between 72 and 94\%, and increases with the pseudoscalar boson mass.

The overlap between the events selected in the four different final states is removed: events that have more isolated muons or electrons than those needed to build the four-lepton final state under study are discarded from the analysis in that final state. Selected leptons are required to be separated from each other by $\Delta R>0.3$, or $>0.4$ if there is a $\tauh$ candidate, since it is built from a jet with a distance parameter of $\Delta R=0.4$.

More than 80\% of the background is rejected by keeping only events for which the visible invariant mass of the four leptons is below 110\GeV in the $\Pgm\Pgm+\Pe\Pgm$ final state, 120\GeV in the $\Pgm\Pgm+\Pe\tauh$ and $\Pgm\Pgm+\Pgm\tauh$ final states, and 130\GeV in the $\Pgm\Pgm+\tauh\tauh$ final state. The threshold depends on the final state because of the different number of neutrinos from $\Pgt$ lepton decays.  Because of the neutrinos, the visible invariant mass is expected to peak below 125\GeV for the signal, and this selection criterion has a signal efficiency close to 100\%. Additionally, the visible mass of the $\Pgt\Pgt$ pair is required to be smaller than the dimuon mass. Events that have a reconstructed dimuon mass lower than 14\GeV or higher than 64\GeV are rejected from the signal region.

The selection described above is optimized for the $\processmmtt$ signal process, which benefits from an excellent dimuon mass resolution of the CMS detector. Assuming a 2HDM+S model, the yield of the $\processtttt$ signal after the selection is between 13 and 52\% of all $\Ph\to\Pa\Pa$ signal events, depending on the final state. The largest fraction is obtained in the $\Pgm\Pgm + \Pe\Pgm$ final state, where the lepton \pt thresholds are the lowest, while the lowest fraction appears in the $\Pgm\Pgm + \tauh\tauh$ final state, which has the highest lepton \pt thresholds.

\section{Estimation of the background with misidentified \texorpdfstring{$\Pgt$}{tau} leptons}
\label{sec:bkg}

The background composed of events where at least one jet is misidentified as one of the final-state leptons is estimated from data. Such events include mostly $\PZ+\text{jets}$ and $\PW\PZ+\text{jets}$ events, but there are also minor contributions from $\PZ\PZ\to 2\ell2$q events, $\ttbar$ production, or from the background from SM events comprised uniquely of jets produced through the strong interaction, referred to as QCD multijet events. The yield and the distributions of these backgrounds are estimated from data via a two-step procedure:
\begin{enumerate}
\item The shape is obtained from data in a signal and $\PZ\PZ$ background free control region with the $\Pgt$ candidates of same sign (SS). To increase the statistical precision of the templates and enrich the region in events with jets misidentified as leptons, the isolation criteria on the $\Pgt$ candidates are relaxed and $\tauh$ candidates are allowed to be also reconstructed as 2-prong decays. Including the 2-prong decays increases the data yield in the control region by about 50\%.
\item The yield is estimated from data events that have one or two nonisolated $\Pgt$ candidates. These events are reweighted with factors that describe the probability for jets to pass the isolation criteria used to select the $\Pgt$ candidates.
The misidentification probabilities for jets are measured in $\PZ\to\Pgm\Pgm+\text{jets}$ events, selected with the same selection criteria as in the signal region except that neither isolation, nor identification criteria are applied to the $\Pgt$ candidates, which are further required to have SS. Additionally the dimuon pair is required to have an invariant mass between 70 and 110\GeV. The probabilities are measured separately in the barrel and in the endcaps as a function of the \pt of the jet that is closest to the lepton, and are parameterized with Landau functions.
\end{enumerate}

The estimation method for the background with jets misidentified as leptons is validated in three control regions: one containing events that pass the full signal selection except that the four-lepton mass criterion is inverted; another where $\tauh$ candidates are reconstructed as 2-prong decays only; and a third one with two SS $\Pgt$ candidates. The background predictions and data are statistically compatible, with deviations not exceeding 20--40\% depending on the final state. The background estimation method has also been validated in simulation for $\PW\PZ+\text{jets}$ and $\PZ+\text{jets}$ events.

\section{Signal and background modeling}

\begin{figure*}[hbpt]
\centering
        \includegraphics[width=0.48\textwidth]{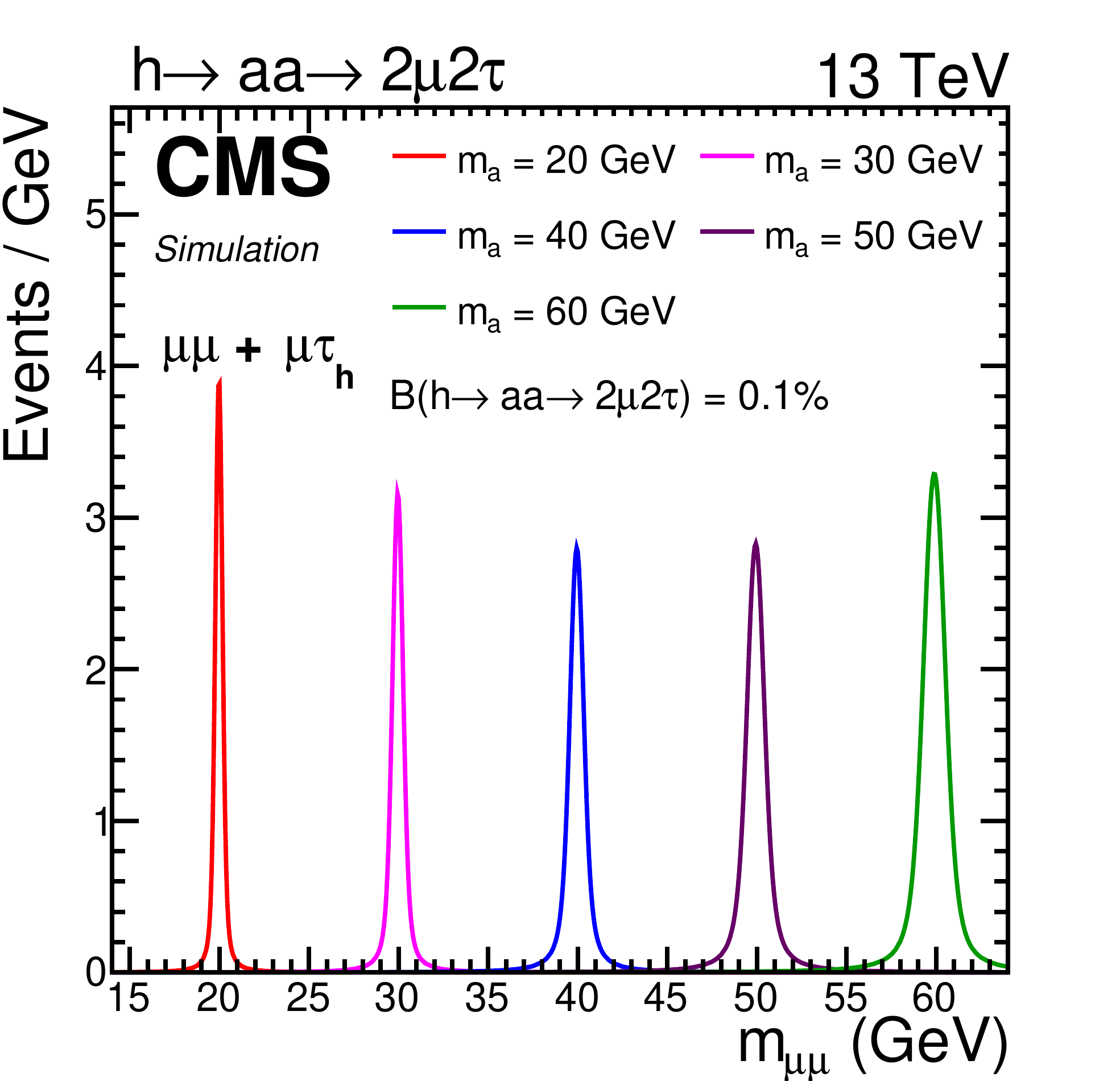}
        \includegraphics[width=0.48\textwidth]{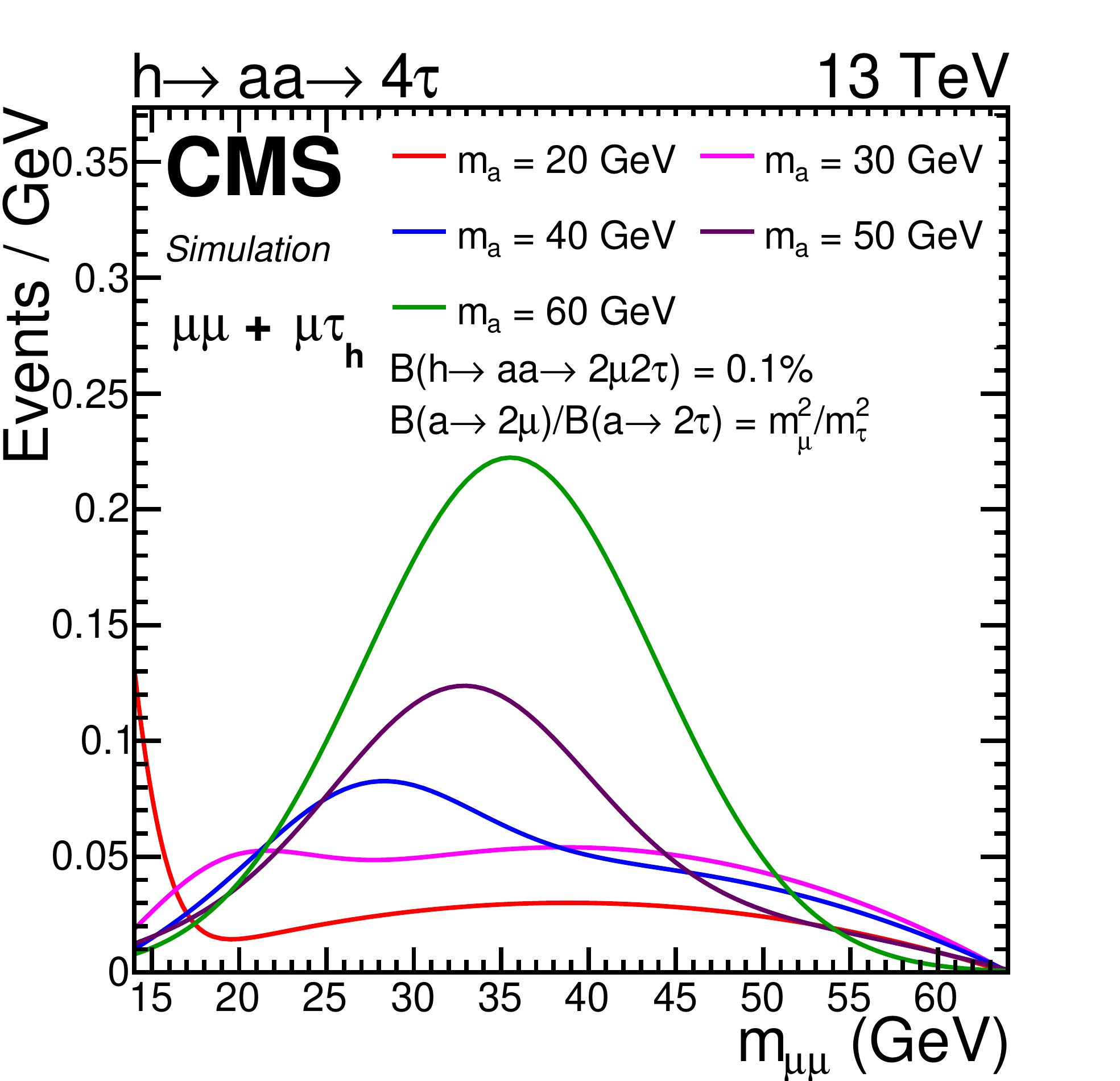}
\caption{Parameterized dimuon invariant mass distributions of the $\processmmtt$ (left) and $\processtttt$ (right) signal processes simulated at different $\ma$ values in the $\Pgm\Pgm+\Pgm\tauh$ final state. The normalization corresponds to the number of expected signal events after the selection for an integrated luminosity of 35.9\fbinv, assuming the production cross section of the Higgs boson predicted in the SM, and $\mathcal{B}(\processmmtt)=2\mathcal{B}(\Ph\to\Pa\Pa)\mathcal{B}(\Pa\to\Pgm\Pgm)\mathcal{B}(\Pa\to\Pgt\Pgt)=0.1\%$. The yield of the $\processtttt$ contribution is further rescaled according to the relation in Eq.~(\ref{eq:mm}). }
\label{fig:signalModel}
\end{figure*}

The results are extracted by fitting the reconstructed dimuon mass distributions. The dimuon mass distributions of the simulated $\processmmtt$ signal events passing all selection criteria are parameterized with Voigt functions, which are convolutions of the Gaussian and Lorentzian profiles with a common mean. The parameterizations for different $\ma$ values in the $\Pgm\Pgm+\Pgm\tauh$ final state are shown in Fig.~\ref{fig:signalModel} (left). The dimuon mass resolution is better than 2\% for all masses and final states considered in the analysis. The parameters of the Voigt functions are fit for each simulated mass and for each final state. The parameters are interpolated for signal masses not covered by simulation.

For the $\processtttt$ signal, the two reconstructed muons that have been chosen to form the dimuon mass distribution can come from either pseudoscalar boson. When the two muons come from the same boson, their visible mass distribution is a wide peak below $\ma$ because they originate from $\Pgt$ lepton decays. When the two muons come from different bosons, they do not form a resonance and their mass distribution is rather flat, with a shape sculpted by kinematic selections. The dimuon mass distribution of the $\processtttt$ signal is parameterized with the sum of a Gaussian function for the resonant contribution and of a polynomial for the nonresonant contribution. The parameterizations for different $\ma$ values in the $\Pgm\Pgm+\Pgm\tauh$ final state are shown in Fig.~\ref{fig:signalModel} (right).

\begin{figure*}[hbpt]
\centering
        \includegraphics[width=0.48\textwidth]{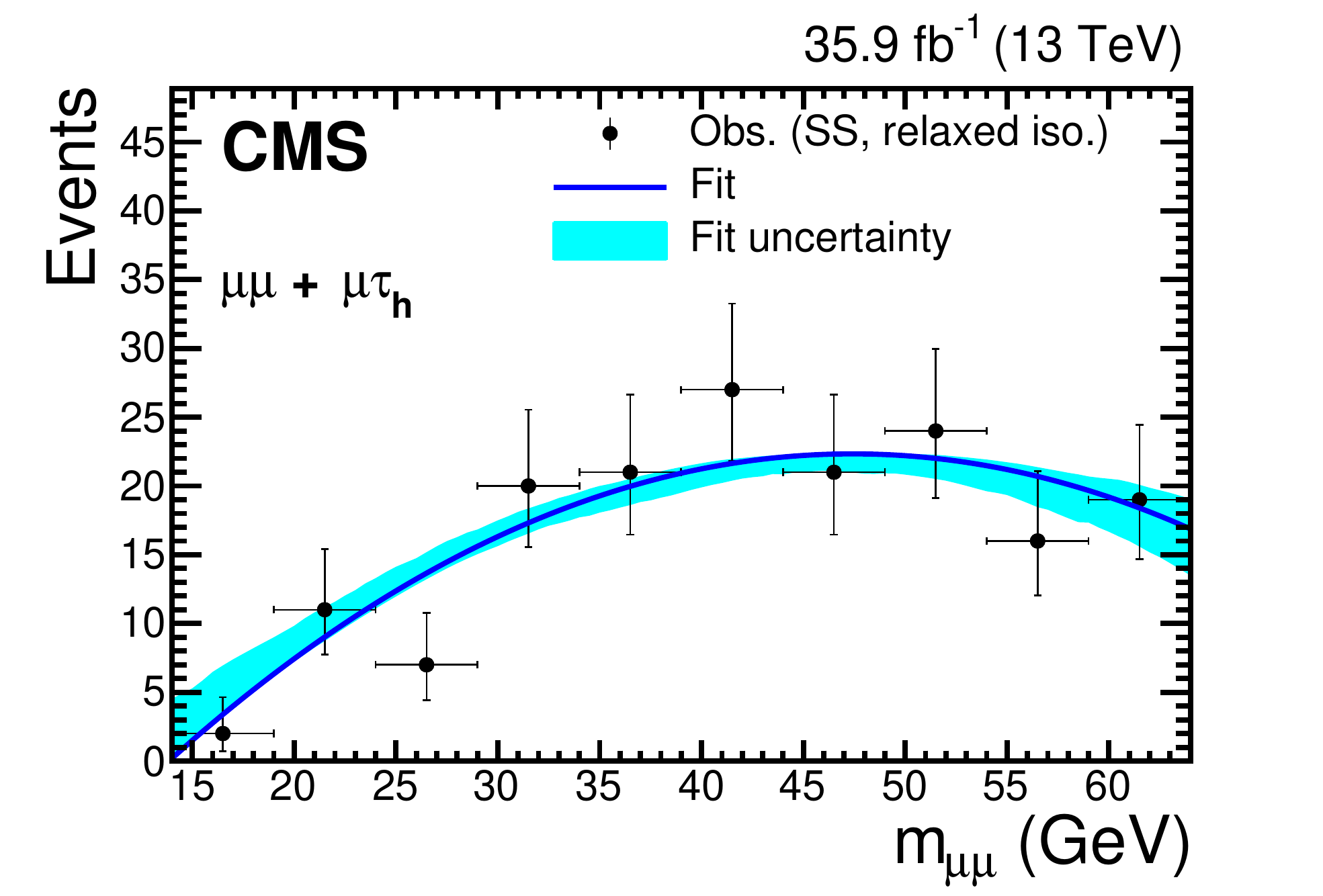}
        \includegraphics[width=0.48\textwidth]{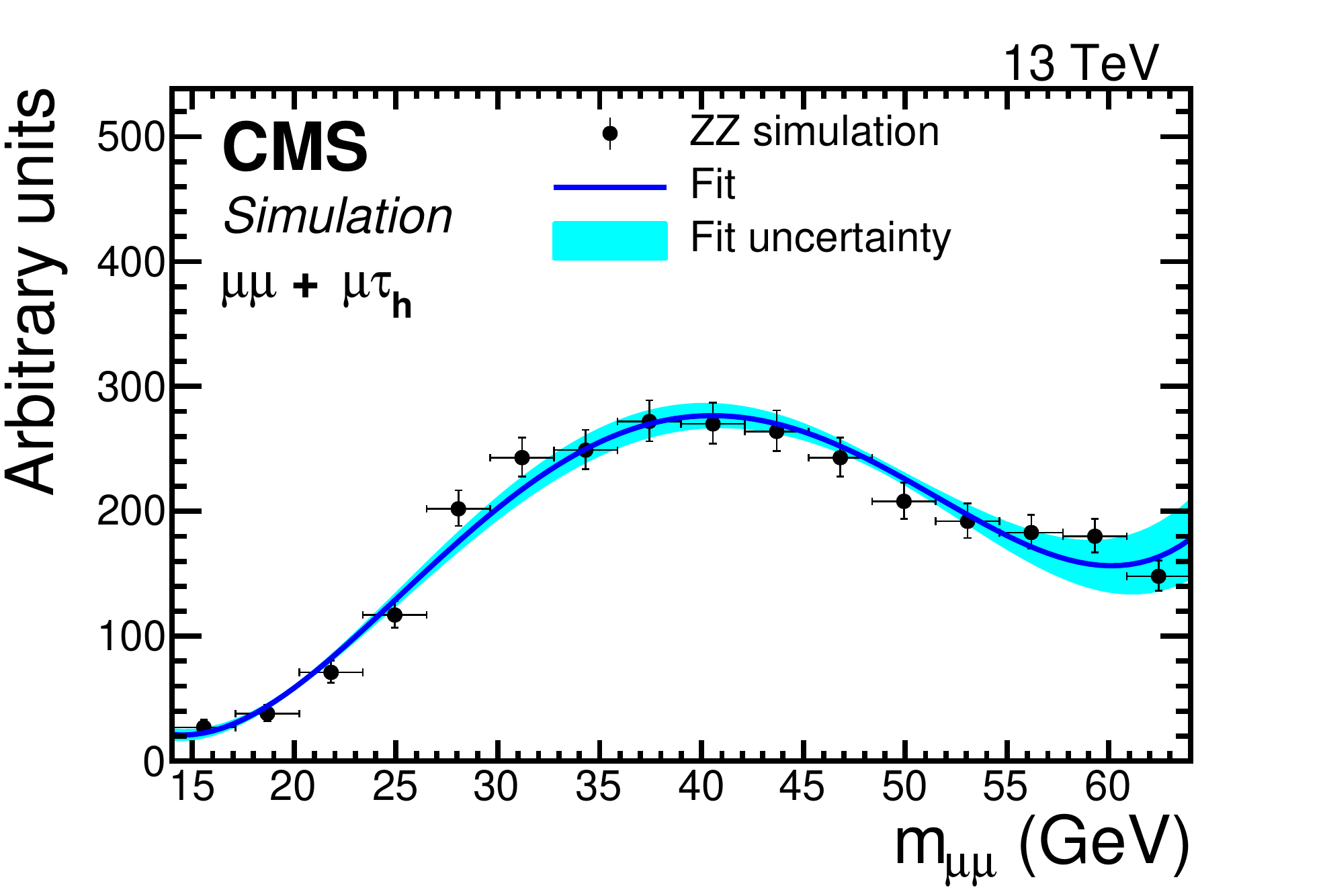}
\caption{Parameterization of the shape of the background with misidentified $\Pgt$ leptons (left) and $\PZ$ pair production background (right) in the $\Pgm\Pgm+\Pgm\tauh$ final state. The points for the $\PZ\PZ$ background represent events selected in simulation, whereas they correspond to observed data events in the SS region with relaxed isolation for the background with misidentified $\Pgt$ leptons.}
\label{fig:bkgModel}
\end{figure*}

The dimuon mass distributions of the $\PZ$ pair background and the background with misidentified $\Pgt$ leptons are parameterized with Bernstein polynomials.  The number of degrees of the polynomial required to describe the background in each channel is determined with a Fisher $F$-test~\cite{ftest}, which selects the minimal number that allows for a good fit quality.  The parameterizations of the backgrounds in the $\Pgm\Pgm+\Pgm\tauh$ final state are shown in Fig.~\ref{fig:bkgModel}. The choice of the fit function and of its degree has only a limited impact on the final results because of the low expected background yields.

\section{Systematic uncertainties}

Yield uncertainties for the processes estimated from simulation include the uncertainty in the integrated luminosity (2.5\%)~\cite{CMS-PAS-LUM-17-001}, in the trigger efficiency (2\%), and in the vetoing of \cPqb-tagged jets (0.5\%). Additionally, the identification, isolation, and reconstruction uncertainties amount to 2\% per muon, 2\% per electron, and 5\% per $\tauh$ candidate. The uncertainty in the $\tauh$ energy scale leads to yield uncertainties between 1 and 2\%. The uncertainty in the yield of the $\PZ\PZ$ background is 12\%: it accounts for the uncertainties in the renormalization and factorization scales, as well as for the uncertainty related to the absence of higher-order electroweak corrections in simulation. The statistical uncertainty related to the limited size of the $\PZ\PZ$ simulated sample reaches up to 13\% in the $\Pgm\Pgm+\tauh\tauh$ final state, but is well below 3\% in the other final states. The uncertainty in the normalization of the signal shapes arising from the parameterization of the normalization as a function of the mass is 5\% per final state. 
The shape uncertainties related to the parameterization of the signal consist of a 0.1\% uncertainty in the mean of the Voigt profile and an anticorrelated 30\% uncertainty in the two width parameters. 

The yield uncertainty in the background with jets misidentified as $\Pgt$ leptons accounts for two different components: the level of agreement between data and background prediction in the control regions, and the statistical uncertainty in the yield predicted in the signal region. As discussed in Section~\ref{sec:bkg}, the first component varies between 20 and 40\%, depending on the final state, whereas the second one ranges between 11 and 23\%. The uncertainties in the parameters of the polynomials used to parameterize the distributions of the background with jets misidentified as $\Pgt$ leptons are included as nuisance parameters in the fit. These parameter uncertainties are obtained from the fits to the data control regions with same sign $\Pgt$ candidates passing relaxed isolation and reconstruction conditions. The uncertainty related to the choice of the fit function for the backgrounds is negligible with respect to the size of the statistical uncertainty. This has been verified by comparing the expected upper limits on the signal when other functional forms are chosen to parameterize the backgrounds.

\section{Results}
\label{sec:results}

\begin{figure*}[hbpt]
\centering
        \includegraphics[width=0.46\textwidth]{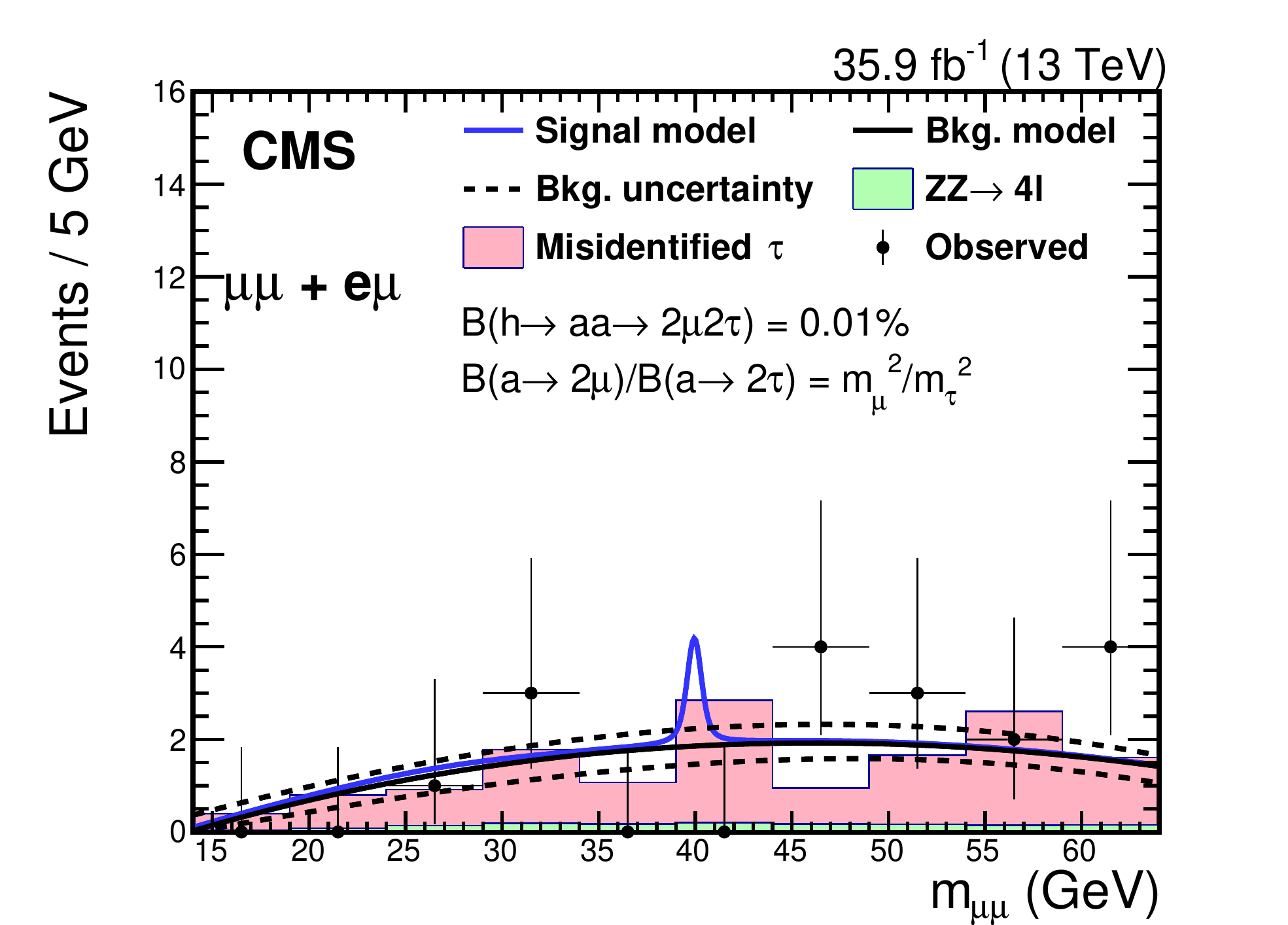}
        \includegraphics[width=0.46\textwidth]{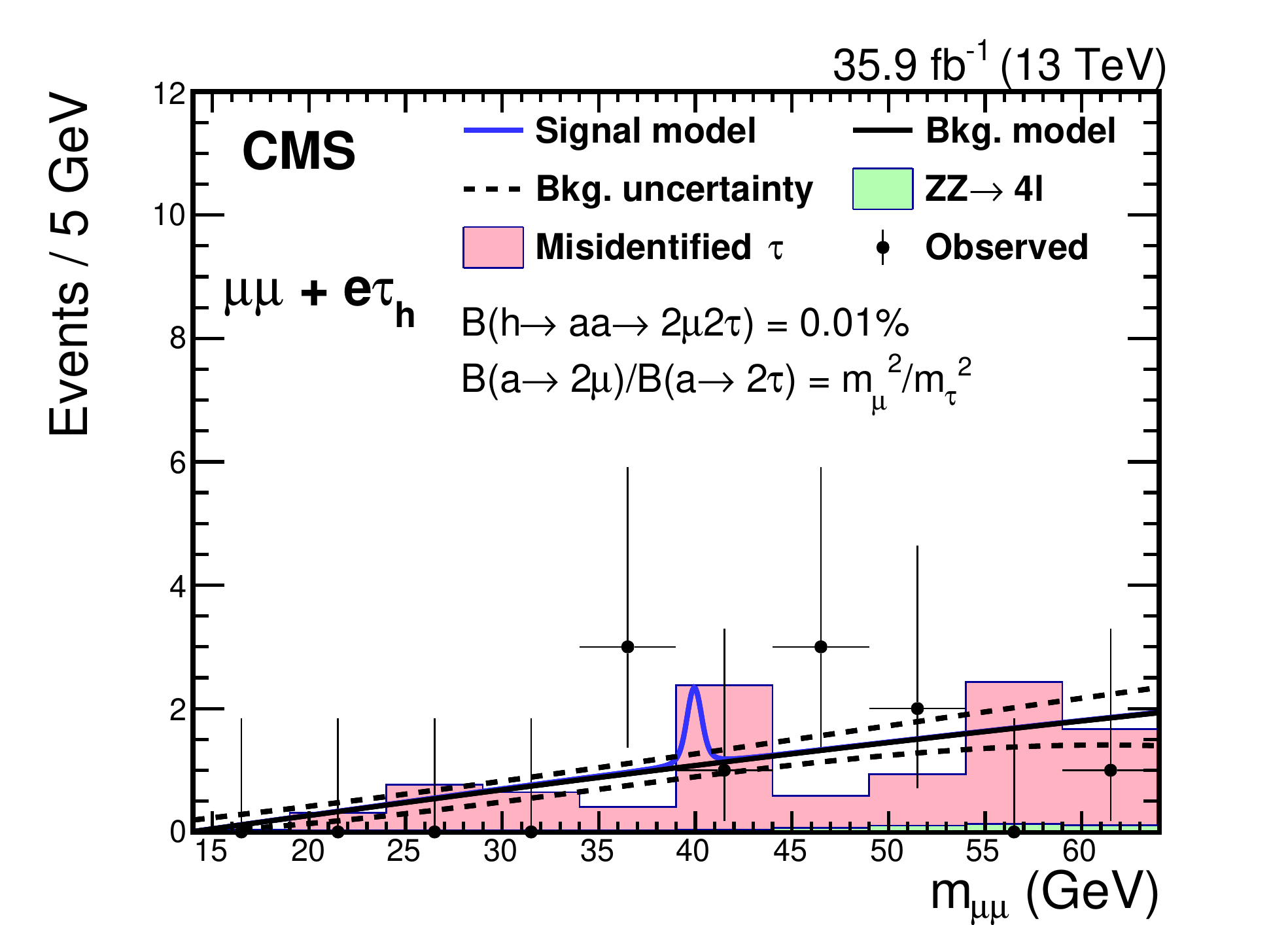}\\
        \includegraphics[width=0.46\textwidth]{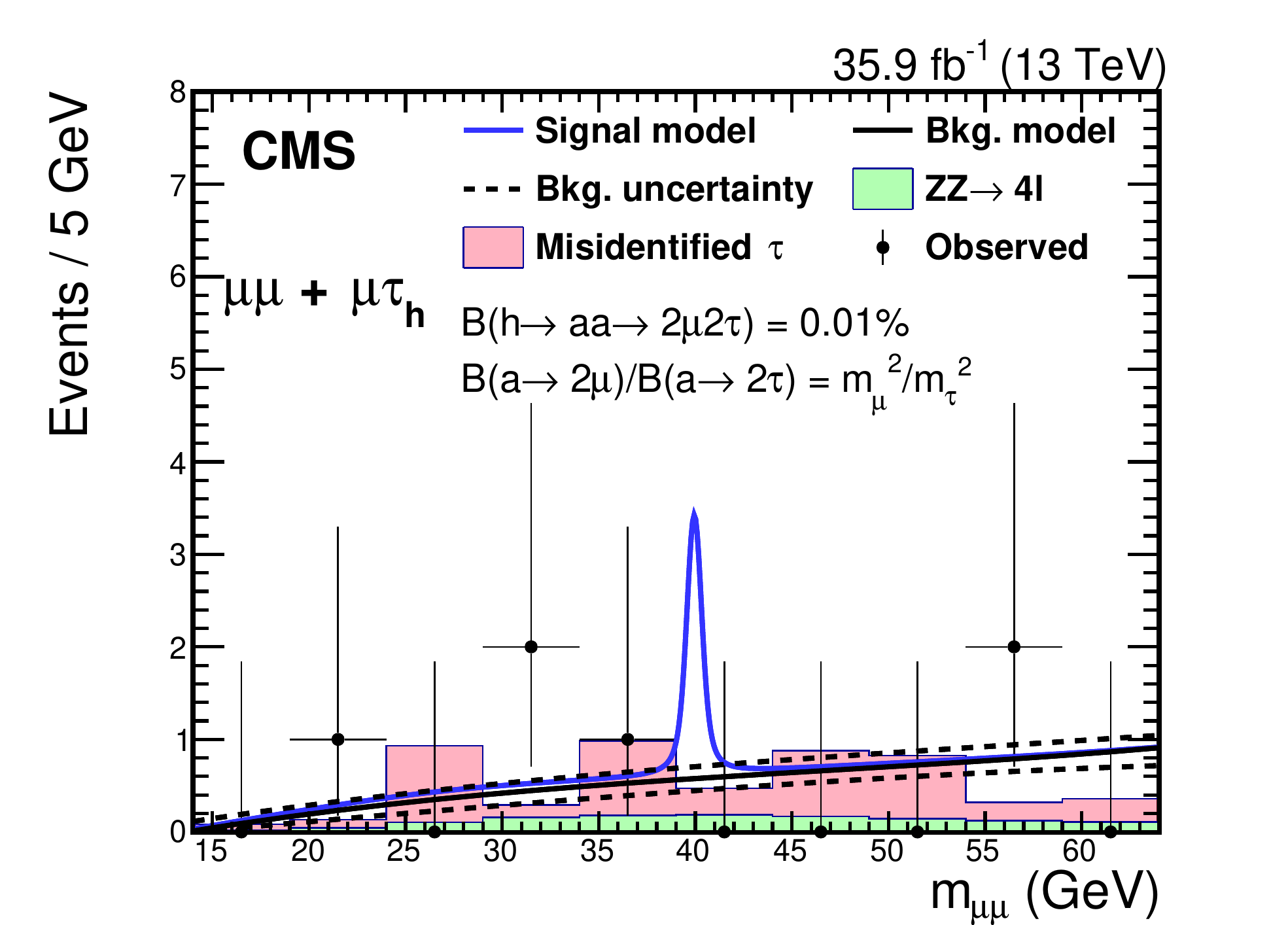}
        \includegraphics[width=0.46\textwidth]{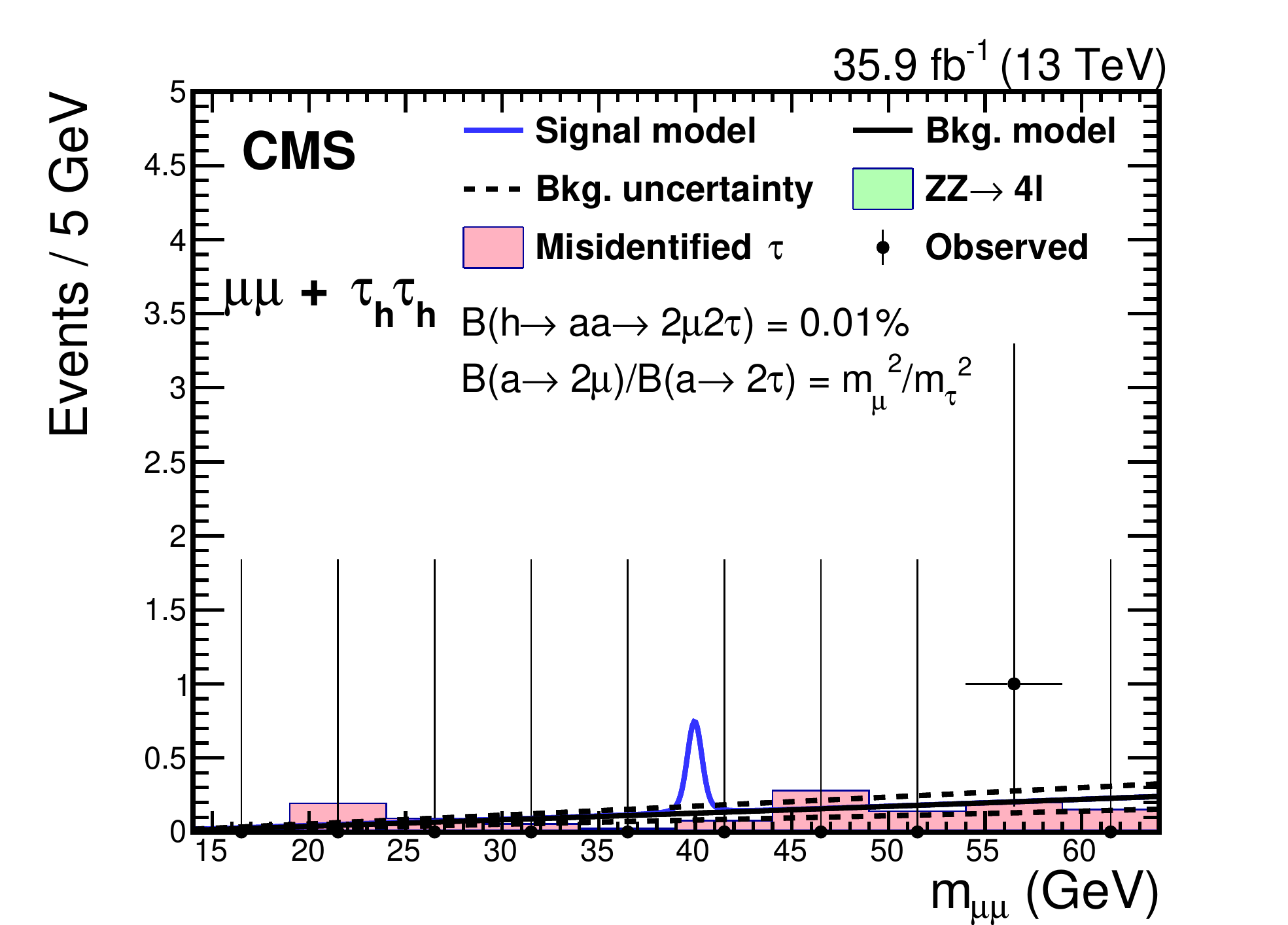}
    \caption{Dimuon mass distributions in the $\Pgm\Pgm+\Pe\Pgm$ (upper left), $\Pgm\Pgm+\Pe\tauh$ (upper right), $\Pgm\Pgm+\Pgm\tauh$ (lower left), and $\Pgm\Pgm+\tauh\tauh$ (lower right) final states. The total background estimate and its uncertainty are given by the black lines. The histograms for the two background components are shown for illustrative purposes only as the background models are extracted from unbinned fits. The signal model is drawn in blue above the background model: it includes both $\processmmtt$ and $\processtttt$, and is normalized using $\mathcal{B}(\processmmtt) = 0.01\%$, assuming the relation in Eq. (\ref{eq:mm}) to determine the relative proportion of these processes. The production cross section of the Higgs boson predicted in the SM is assumed. }
    \label{fig:cmb}
\end{figure*}

To test for the existence of a resonance, an unbinned maximum-likelihood fit to the dimuon invariant mass distribution is performed. In the fit, the systematic uncertainties are nuisance parameters varied according to a log-normal probability density function for the yield uncertainties and a Gaussian probability density function for the shape uncertainties. The dimuon mass distributions for the four final states are shown in Fig.~\ref{fig:cmb}. The expected background and signal yields in the signal region are given in Table \ref{tab:yields} for the four final states.

\begin{table*}
\topcaption{Yields of the signal and background processes in the four final states, as well as the number of observed events in each final state, in the dimuon mass range between 14 and 64\GeV. The signal yields are given for $\mathcal{B}(\processmmtt) = 0.01\%$. The $\processtttt$ signal is scaled assuming the couplings of the pseudoscalar boson proportional to the squared lepton mass, as in Eq. (\ref{eq:mm}). The production cross section of the Higgs boson predicted in the SM is assumed. The uncertainties combine the statistical and systematic sources.}

\label{tab:yields}
\centering
\begin{tabular}{lcccc}
 & $\Pgm\Pgm+\Pe\Pgm$ & $\Pgm\Pgm + \Pe\tauh$ & $\Pgm\Pgm+\Pgm\tauh$ & $\Pgm\Pgm+\tauh\tauh$ \\
\hline
$\PZ\PZ\to 4\ell$ & $1.5\pm 0.2$ & $0.5\pm0.1$ & $1.2\pm0.2$ & $0.03\pm0.01$ \\
Misidentified $\Pgt$ & $13.2\pm5.5$ & $9.7\pm2.5$ & $4.0\pm1.2$ & $1.2\pm0.5$ \\[\cmsTabSkip]
$\processmmtt$, $\ma=20\GeV$ & $0.39$ & $0.25$ & $0.47$ & $0.10$ \\
$\processtttt$, $\ma=20\GeV$ & $0.37$ & $0.04$ & $0.24$ & $0.01$ \\
$\processmmtt$, $\ma=40\GeV$ & $0.57$ & $0.28$ & $0.68$ & $0.14$ \\
$\processtttt$, $\ma=40\GeV$ & $0.68$ & $0.09$ & $0.48$ & $0.02$ \\
$\processmmtt$, $\ma=60\GeV$ & $0.94$ & $0.85$ & $1.18$ & $0.52$ \\
$\processtttt$, $\ma=60\GeV$ & $1.27$ & $0.20$ & $0.93$ & $0.05$ \\[\cmsTabSkip]
Observed & 17 & 10 & 6 & 1 \\
\end{tabular}
\end{table*}

\begin{figure*}[hbpt]
\centering
        \includegraphics[width=0.36\textwidth]{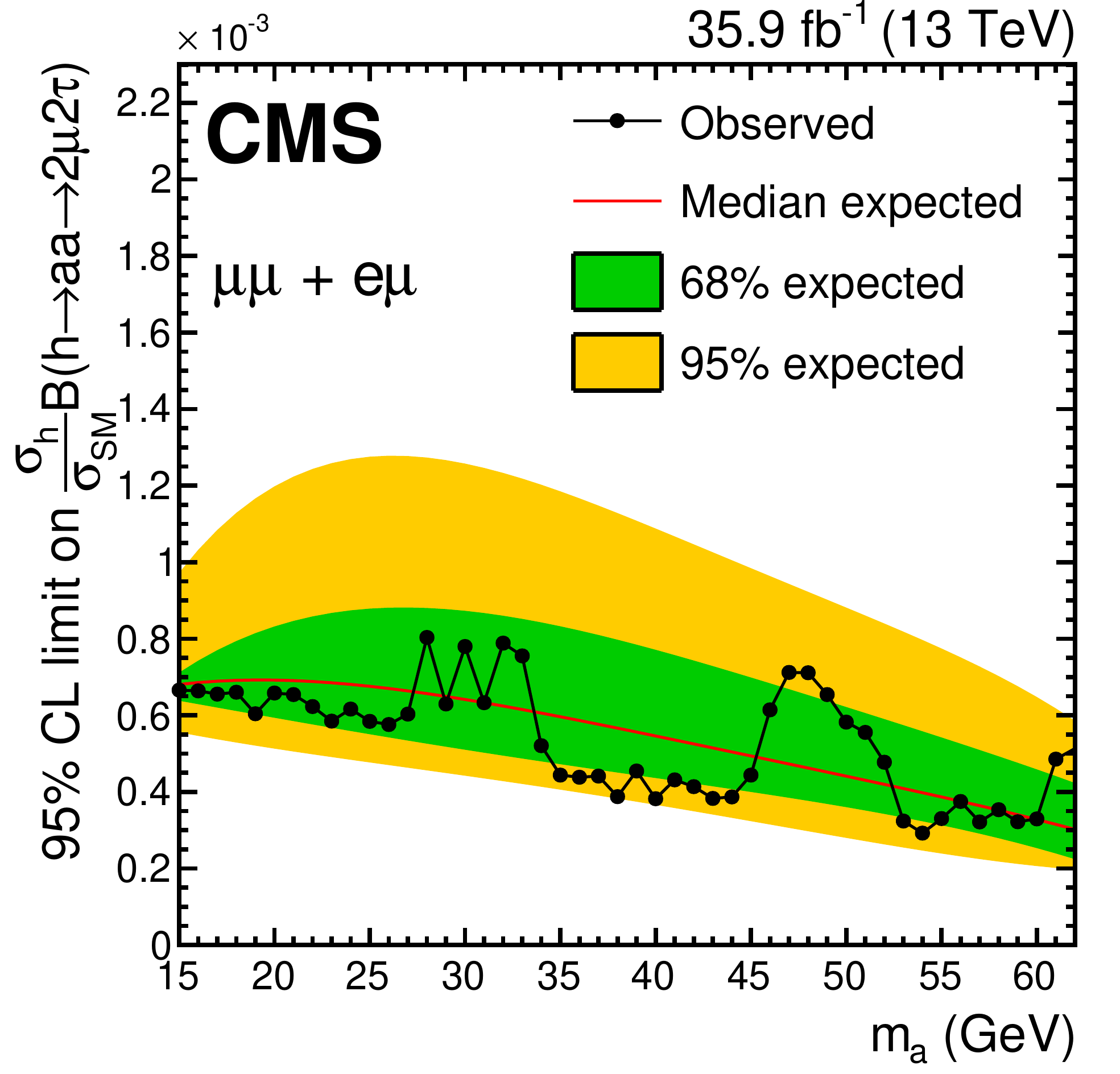}
        \includegraphics[width=0.36\textwidth]{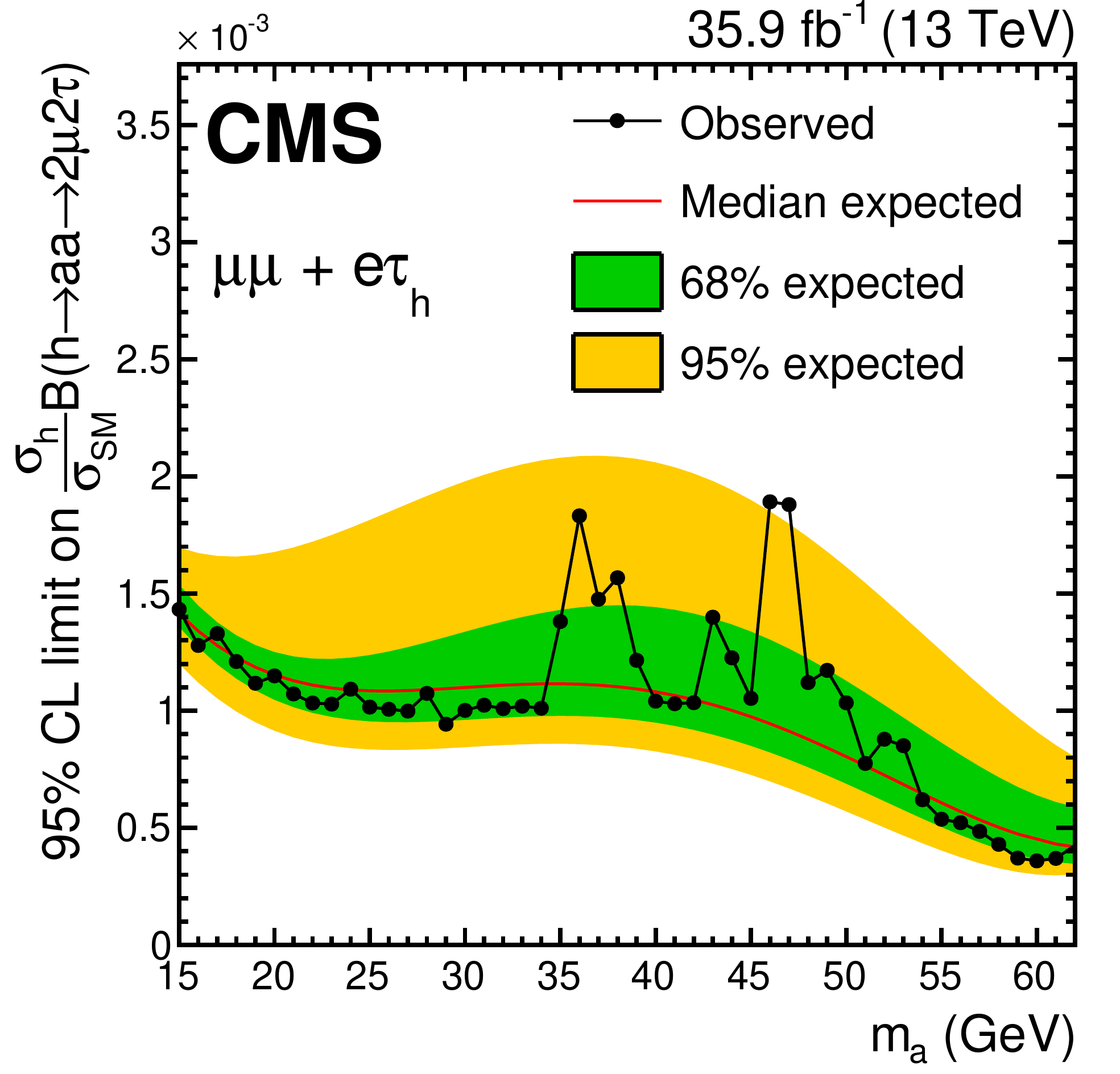}\\
        \includegraphics[width=0.36\textwidth]{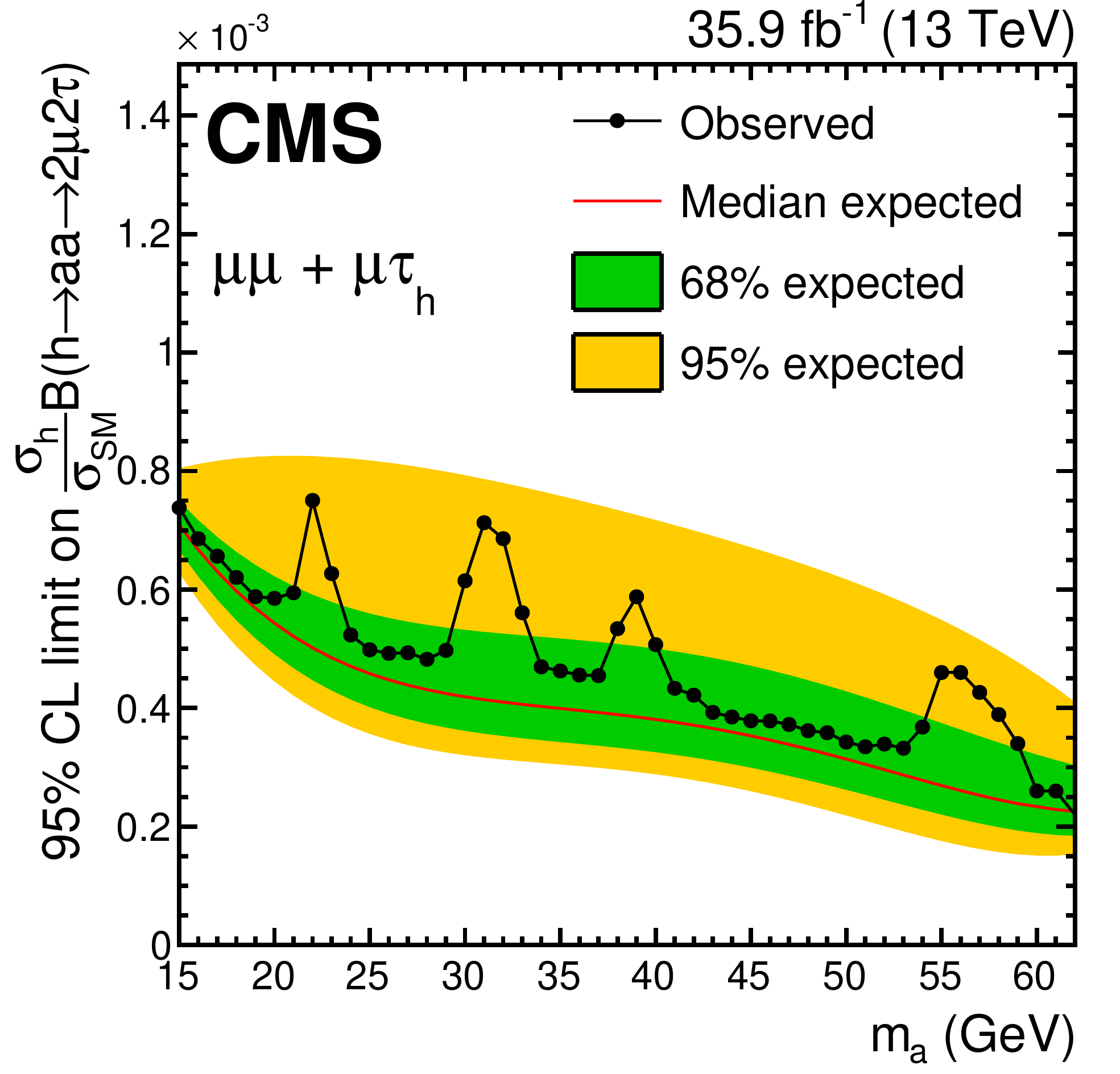}
        \includegraphics[width=0.36\textwidth]{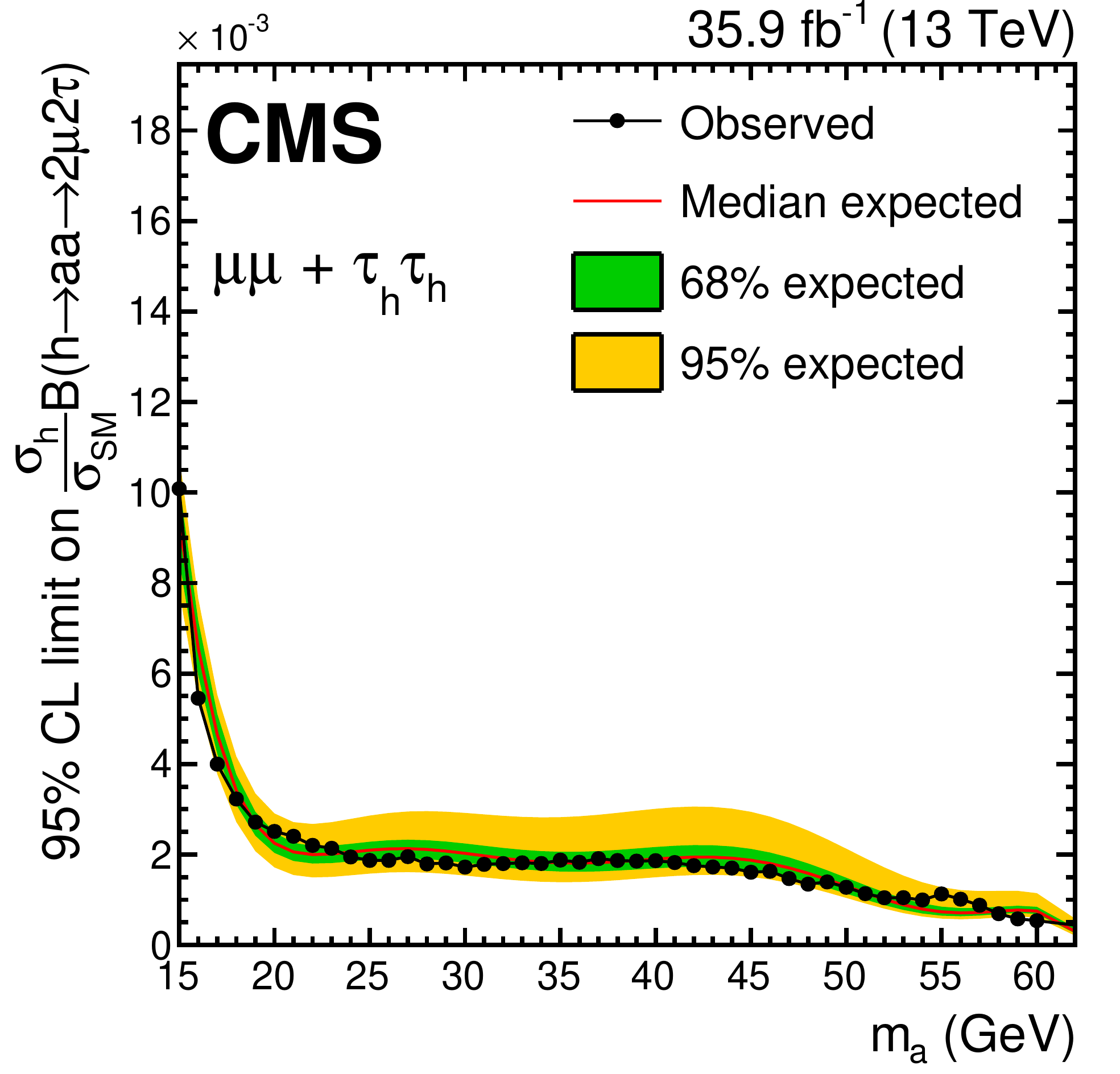}\\
        \includegraphics[width=0.36\textwidth]{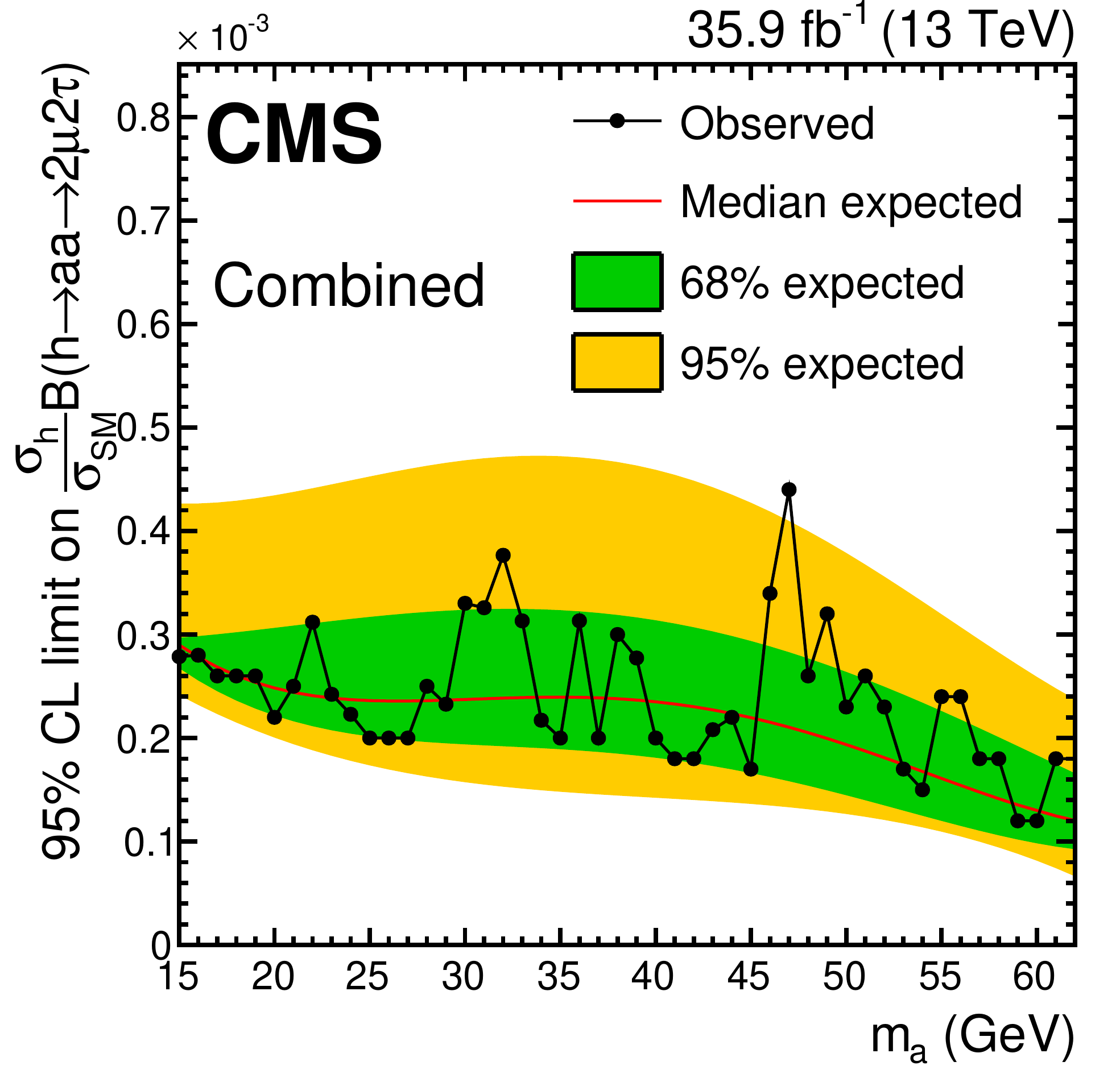}
    \caption{Upper limits at 95\% \CL on $(\sigma_\Ph/\sigma_{\textrm{SM}})\mathcal{B}(\processmmtt)$, in the $\Pgm\Pgm+\Pe\Pgm$ (upper left), $\Pgm\Pgm+\Pe\tauh$ (upper right), $\Pgm\Pgm+\Pgm\tauh$ (middle left), $\Pgm\Pgm+\tauh\tauh$ (middle right) final states, and for the combination of these final states (lower). The $\processtttt$ process is considered as a part of the signal, and is scaled with respect to the $\processmmtt$ signal using Eq. (\ref{eq:mm}).}
    \label{fig:limits}
\end{figure*}

No significant excess of data is observed above the expected SM background. Upper limits at 95\% \CL are set on $(\sigma_\Ph/\sigma_\textrm{SM}) \mathcal{B}(\processmmtt)=2(\sigma_\Ph/\sigma_\textrm{SM})\mathcal{B}(\Ph\to\Pa\Pa)\mathcal{B}(\Pa\to\Pgm\Pgm)\mathcal{B}(\Pa\to\Pgt\Pgt)$ using the modified frequentist construction \CLs~\cite{LHC-HCG-Report,Chatrchyan:2012tx,Junk,Read:2002hq} for pseudoscalar masses between 15.0 and 62.5\GeV. In this expression, $\sigma_\Ph/\sigma_\textrm{SM}$ is the Higgs boson cross section for the gluon fusion and vector boson fusion production modes, divided by its SM prediction. The limits are shown in Fig.~\ref{fig:limits} for the individual final states and for their combination. The combined upper limits on the branching fraction $\mathcal{B}(\Ph\to\Pa\Pa\to2\Pgm2\Pgt)$ are as low as $1.2\times 10^{-4}$ for a mass of 60\GeV assuming the SM production cross section for the Higgs boson.
The expected limits are the tightest for the $\Pgm\Pgm+\Pgm\tauh$ final state because the lepton \pt thresholds are lower than in the $\Pgm\Pgm + \Pe\tauh$ and $\Pgm\Pgm+\tauh\tauh$ final states, and because the branching fraction is larger than in the $\Pgm\Pgm+\Pe\Pgm$ final state. The  $\processtttt$ signal is assumed to scale according to Eq. (\ref{eq:mm}) with respect to the $\processmmtt$ signal. Alternatively, considering a null contribution from $\processtttt$, there is still no significant excess of data over the expected SM background and the expected limits become less stringent by approximately $10\%$.

\begin{figure*}[hbpt]
\centering
    \includegraphics[width=0.46\textwidth]{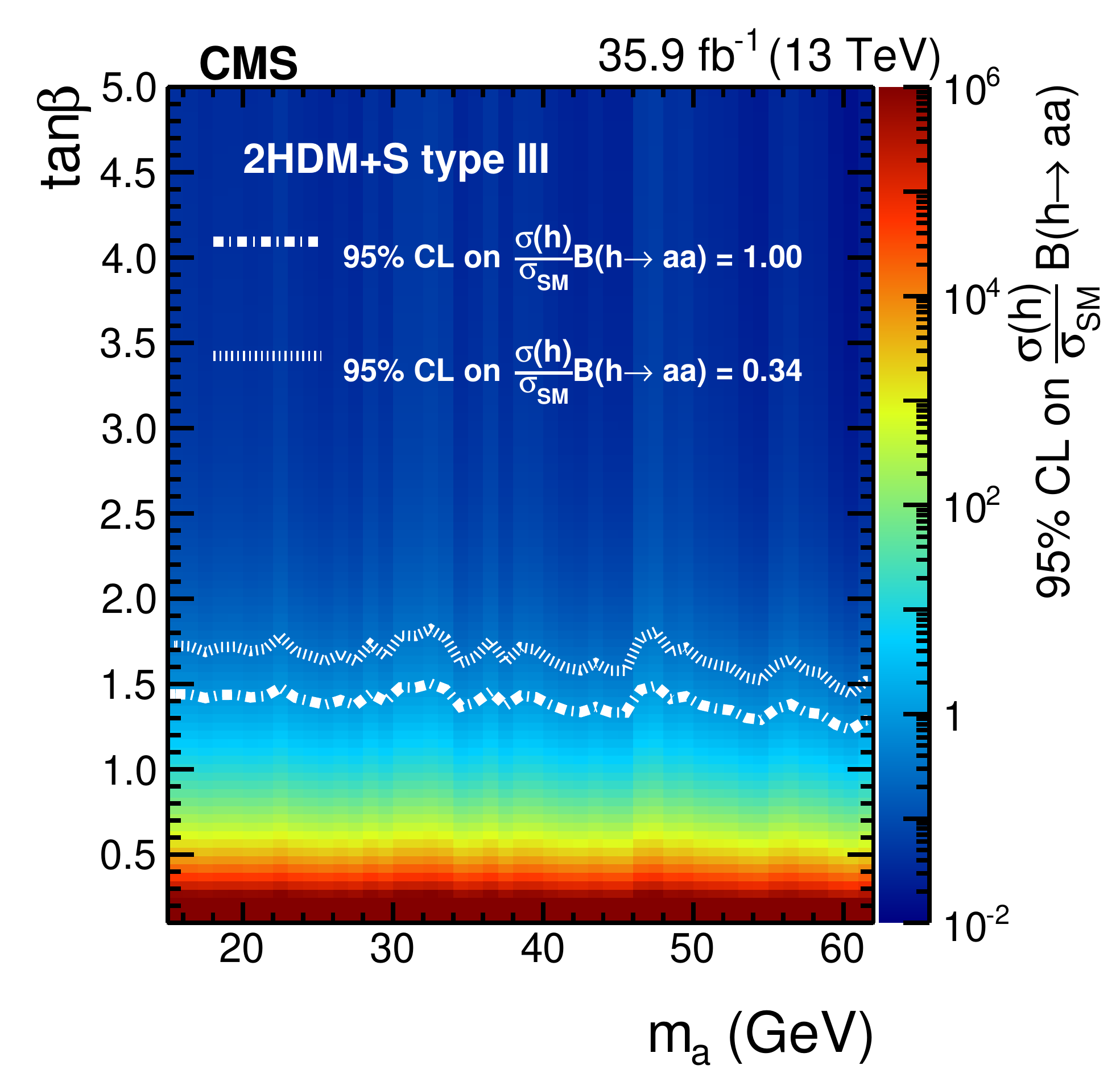}
    \includegraphics[width=0.46\textwidth]{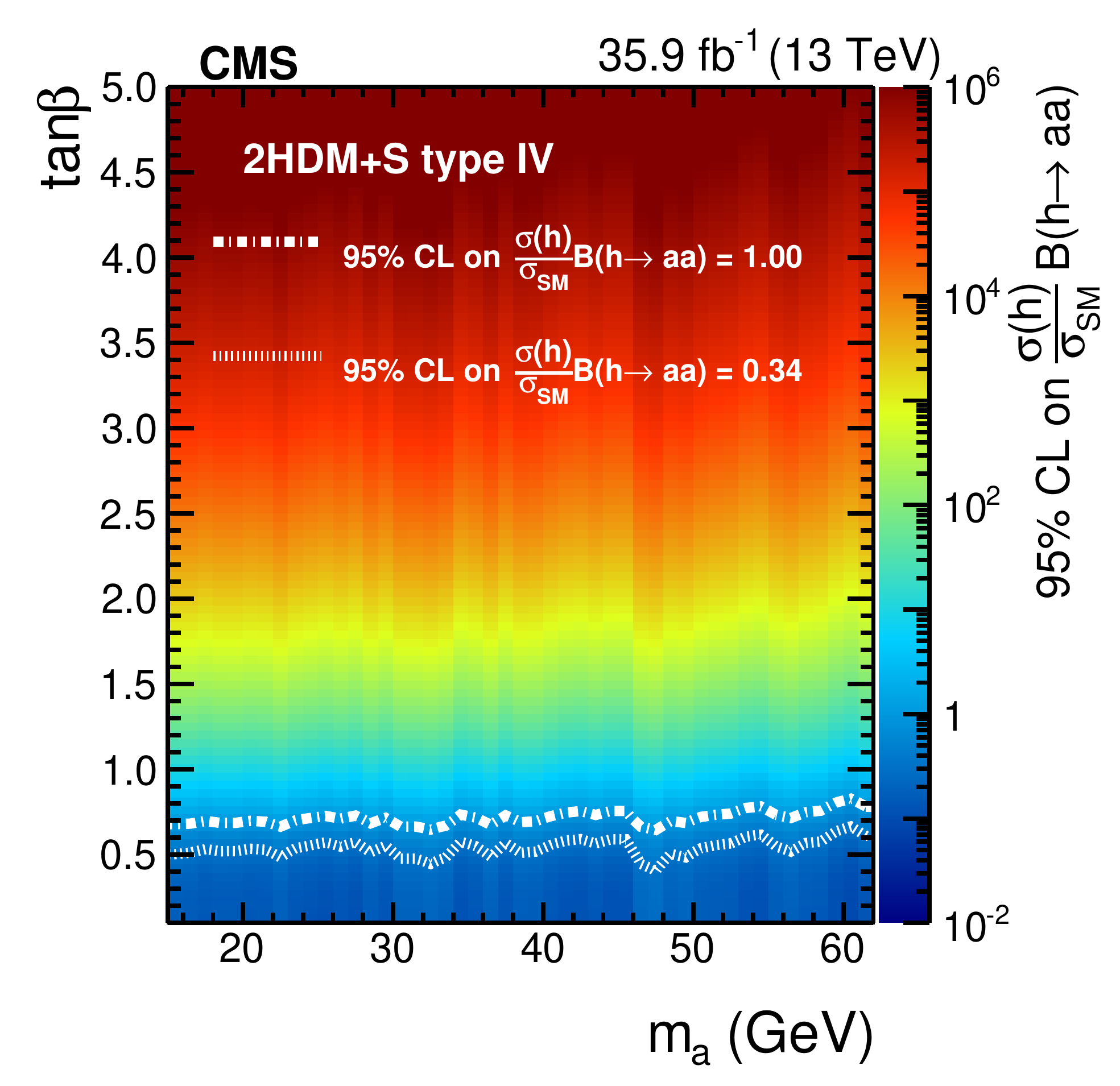}
    \caption{Observed limits on $(\sigma_{\Ph}/\sigma_{\mathrm{SM}})\mathcal{B}(\Ph\to \Pa\Pa)$ in 2HDM+S type III (left) and type IV (right). The contour lines shown for $\mathcal{B}(\Ph\to \Pa\Pa)=1.0$ and 0.34 correspond to the colour scale indicated on the right vertical scale. The number 0.34 corresponds to the limit on the branching fraction of the Higgs boson to beyond-the-SM particles at 95\% \CL obtained with data collected at center-of-mass energies of 7 and 8\TeV by the CMS and ATLAS experiments~\cite{Khachatryan:2016vau}.}
    \label{fig:2hdms}
\end{figure*}

The results can be interpreted as upper limits on $(\sigma_{\Ph}/\sigma_{\textrm{SM}})\mathcal{B}(\Ph\to \Pa\Pa)$ in the different 2HDM+S models.
Types I--IV 2HDM+S forbid flavor changing neutral currents at tree level. In type I 2HDM+S, all SM particles couple to the first doublet and the branching fractions of the light pseudoscalar to SM particles are independent of $\tan\beta$, defined as the ratio of the vacuum expectation value of the second doublet to that of the first doublet. In type II 2HDM+S, including the NMSSM, up-type quarks couple to the first doublet, and leptons and down-type quarks couple to the second doublet. This leads to pseudoscalar decays to leptons and down-type fermions enhanced for $\tan\beta>1$. In these two types, the analysis is sensitive to a cross section larger than approximately three times the SM production cross section of the Higgs boson for $\mathcal{B}(\processmmtt)=100\%$. In type III 2HDM+S, quarks couple to the first doublet and leptons to the second one, making it the most favorable type of 2HDM+S for $\processmmtt$ decays at large $\tan\beta$. In type IV 2HDM+S, leptons and up-type quarks couple to the first doublet while down-type quarks couple to the second doublet. With $\ma$, $\tan\beta$, and the type of 2HDM+S specified, the branching fractions of the pseudoscalars to SM particles can be predicted following the prescriptions in Refs.~\cite{PhysRevD.90.075004,anatomy}. The results expressed as limits on $(\sigma_{\Ph}/\sigma_{\textrm{SM}})\mathcal{B}(\Ph\to \Pa\Pa)$ are shown in Fig.~\ref{fig:2hdms} for the last two types of 2HDM+S. The most stringent limits are obtained in 2HDM+S type III at large $\tan\beta$, where the couplings to leptons are enhanced, and where limits of approximately 3\% are set for $\tan\beta\gtrsim 3$. This analysis improves previous results~\cite{Khachatryan:2017mnf} in the $2\Pgm2\Pgt$ final state by a factor two or more for $15.0<\ma<62.5\GeV$ in all four types of 2HDM+S.

\section{Summary}
\label{sec:summary}

A search for an exotic decay of the Higgs boson to a pair of light pseudoscalars in the final state of two muons and two $\Pgt$ leptons has been performed using data collected by the CMS experiment in 2016 at a center-of-mass energy of 13\TeV, and corresponding to an integrated luminosity of 35.9\fbinv. The results are extracted from an unbinned fit of the dimuon mass spectrum. Limits are set at 95\% confidence level on the branching fraction $\mathcal{B}(\Ph\to\Pa\Pa\to2\Pgm2\Pgt)$ for the masses of the light pseudoscalar between 15.0 and 62.5\GeV, and are as low as $1.2\times 10^{-4}$ for a mass of 60\GeV assuming the SM production cross section for the Higgs boson. These are the most stringent limits obtained in the final state of two muons and two $\Pgt$ leptons for the masses above 15\GeV, improving previous limits~\cite{Khachatryan:2017mnf,Aad:2015oqa} by more than a factor two. They provide the tightest constraints in this mass range on exotic Higgs boson decays in scenarios where the decays of pseudoscalar bosons to leptons are enhanced.

\begin{acknowledgments}
We congratulate our colleagues in the CERN accelerator departments for the excellent performance of the LHC and thank the technical and administrative staffs at CERN and at other CMS institutes for their contributions to the success of the CMS effort. In addition, we gratefully acknowledge the computing centers and personnel of the Worldwide LHC Computing Grid for delivering so effectively the computing infrastructure essential to our analyses. Finally, we acknowledge the enduring support for the construction and operation of the LHC and the CMS detector provided by the following funding agencies: BMWFW and FWF (Austria); FNRS and FWO (Belgium); CNPq, CAPES, FAPERJ, and FAPESP (Brazil); MES (Bulgaria); CERN; CAS, MoST, and NSFC (China); COLCIENCIAS (Colombia); MSES and CSF (Croatia); RPF (Cyprus); SENESCYT (Ecuador); MoER, ERC IUT, and ERDF (Estonia); Academy of Finland, MEC, and HIP (Finland); CEA and CNRS/IN2P3 (France); BMBF, DFG, and HGF (Germany); GSRT (Greece); NKFIA (Hungary); DAE and DST (India); IPM (Iran); SFI (Ireland); INFN (Italy); MSIP and NRF (Republic of Korea); LAS (Lithuania); MOE and UM (Malaysia); BUAP, CINVESTAV, CONACYT, LNS, SEP, and UASLP-FAI (Mexico); MBIE (New Zealand); PAEC (Pakistan); MSHE and NSC (Poland); FCT (Portugal); JINR (Dubna); MON, RosAtom, RAS and RFBR (Russia); MESTD (Serbia); SEIDI, CPAN, PCTI and FEDER (Spain); Swiss Funding Agencies (Switzerland); MST (Taipei); ThEPCenter, IPST, STAR, and NSTDA (Thailand); TUBITAK and TAEK (Turkey); NASU and SFFR (Ukraine); STFC (United Kingdom); DOE and NSF (USA).

\hyphenation{Rachada-pisek} Individuals have received support from the Marie-Curie program and the European Research Council and Horizon 2020 Grant, contract No. 675440 (European Union); the Leventis Foundation; the A. P. Sloan Foundation; the Alexander von Humboldt Foundation; the Belgian Federal Science Policy Office; the Fonds pour la Formation \`a la Recherche dans l'Industrie et dans l'Agriculture (FRIA-Belgium); the Agentschap voor Innovatie door Wetenschap en Technologie (IWT-Belgium); the F.R.S.-FNRS and FWO (Belgium) under the ``Excellence of Science - EOS" - be.h project n. 30820817; the Ministry of Education, Youth and Sports (MEYS) of the Czech Republic; the Lend\"ulet (``Momentum") Programme and the J\'anos Bolyai Research Scholarship of the Hungarian Academy of Sciences, the New National Excellence Program \'UNKP, the NKFIA research grants 123842, 123959, 124845, 124850 and 125105 (Hungary); the Council of Science and Industrial Research, India; the HOMING PLUS program of the Foundation for Polish Science, cofinanced from European Union, Regional Development Fund, the Mobility Plus program of the Ministry of Science and Higher Education, the National Science Center (Poland), contracts Harmonia 2014/14/M/ST2/00428, Opus 2014/13/B/ST2/02543, 2014/15/B/ST2/03998, and 2015/19/B/ST2/02861, Sonata-bis 2012/07/E/ST2/01406; the National Priorities Research Program by Qatar National Research Fund; the Programa Estatal de Fomento de la Investigaci{\'o}n Cient{\'i}fica y T{\'e}cnica de Excelencia Mar\'{\i}a de Maeztu, grant MDM-2015-0509 and the Programa Severo Ochoa del Principado de Asturias; the Thalis and Aristeia programs cofinanced by EU-ESF and the Greek NSRF; the Rachadapisek Sompot Fund for Postdoctoral Fellowship, Chulalongkorn University and the Chulalongkorn Academic into Its 2nd Century Project Advancement Project (Thailand); the Welch Foundation, contract C-1845; and the Weston Havens Foundation (USA). \end{acknowledgments}

\bibliography{auto_generated}

\cleardoublepage \appendix\section{The CMS Collaboration \label{app:collab}}\begin{sloppypar}\hyphenpenalty=5000\widowpenalty=500\clubpenalty=5000\vskip\cmsinstskip
\textbf{Yerevan Physics Institute, Yerevan, Armenia}\\*[0pt]
A.M.~Sirunyan, A.~Tumasyan
\vskip\cmsinstskip
\textbf{Institut f\"{u}r Hochenergiephysik, Wien, Austria}\\*[0pt]
W.~Adam, F.~Ambrogi, E.~Asilar, T.~Bergauer, J.~Brandstetter, M.~Dragicevic, J.~Er\"{o}, A.~Escalante~Del~Valle, M.~Flechl, R.~Fr\"{u}hwirth\cmsAuthorMark{1}, V.M.~Ghete, J.~Hrubec, M.~Jeitler\cmsAuthorMark{1}, N.~Krammer, I.~Kr\"{a}tschmer, D.~Liko, T.~Madlener, I.~Mikulec, N.~Rad, H.~Rohringer, J.~Schieck\cmsAuthorMark{1}, R.~Sch\"{o}fbeck, M.~Spanring, D.~Spitzbart, A.~Taurok, W.~Waltenberger, J.~Wittmann, C.-E.~Wulz\cmsAuthorMark{1}, M.~Zarucki
% ...
\vskip\cmsinstskip
% ...
1:  Also at Vienna University of Technology, Vienna, Austria\\
% ...
\end{sloppypar}
\end{document}